\documentclass[%
 reprint,
 amsmath,amssymb,
 aps,
]{revtex4-1}

\usepackage{graphicx}
\usepackage{dcolumn}
\usepackage{bm}
\usepackage{placeins}
\usepackage{xcolor}

\begin{document}

\preprint{APS/123-QED}

\title{Transverse Beam Stability with Low-Impedance Collimators in the\\ High Luminosity Large Hadron Collider: Status and Challenges}

\author{S.~A.~Antipov}
 \email{Sergey.Antipov@cern.ch}%
\author{D.~Amorim}%
\author{A.~Bertarelli}
\author{N.~Biancacci}%
\author{R.~Bruce}
\author{E.~Carideo}
\author{F.~Carra}
\author{J.~Guardia~Valenzuela}
\author{A.~Mereghetti}%
\author{E.~M\'etral}%
\author{B.~Salvant}
\author{S.~Redaelli}
\author{D.~Valuch}

\affiliation{%
 CERN, Geneva 1211, Switzerland
}%

\date{\today}

\begin{abstract}
The High-Luminosity LHC Project aims to increase the integrated luminosity that will be collected by the Large Hadron Collider for the needs of the high energy physics frontier by the end of its Run 3 by more than a factor ten. This will require doubling the beam intensity, and in order to ensure coherent stability until the brighter beams are put in collision, the transverse impedance of the machine has to be reduced. As the major portion of the ring impedance is generated by its collimation system, several low resistivity jaw materials have been considered to lower the collimator impedance and a special collimator has been built and installed in the machine to study their effect. In order to assess the performance of each material we performed a series of tune shift measurements with LHC beams. The results show a significant reduction of the resistive wall tune shift with novel materials, in good agreement with the impedance model and the bench impedance and resistivity measurements. The largest improvement is obtained with a molybdenum coating of a molybdenum-graphite jaw. This coating, applied to the most critical collimators, is estimated to lower the machine impedance by up to 30\% and the stabilizing Landau octupole threshold by up to 240~A. A half of the overall improvement can be obtained by coating the jaws of a subset of 4 out of 11 collimators identified as the highest contributors to machine impedance. This subset of low-impedance collimators is being installed during the Long Shutdown 2 in 2019-2020.
\end{abstract}

\pacs{Valid PACS appear here}
\maketitle


\section{\label{sec:Intro}LHC collimation system and transverse beam stability}

Collimators are widely used in particle accelerators. The systems find their usage from linear coherent light sources such as Linear Coherent Light Source \citep{bib:LCLS}, SwissFEL \citep{Bettoni:2018env}, or Next Generation Light Source \citep{bib:NGLS} to high-intensity circular colliders, both past, present, and proposed: Tevatron \citep{bib:TeV}, Relativistic Heavy Ion Collider \citep{bib:RHIC}, Large Hadron Collider (LHC) \citep{bib:w1}, or Future Circular Collider (FCC) \citep{bib:FCC-hh}. In superconducting colliders collimators play an essential role of protecting their superconducting magnets from quenches or damage in case of beam losses as well as controlling the beam halo \citep{bib:w2}.

\begin{figure}[b]
  \centering
  \includegraphics[width = 3.375in]{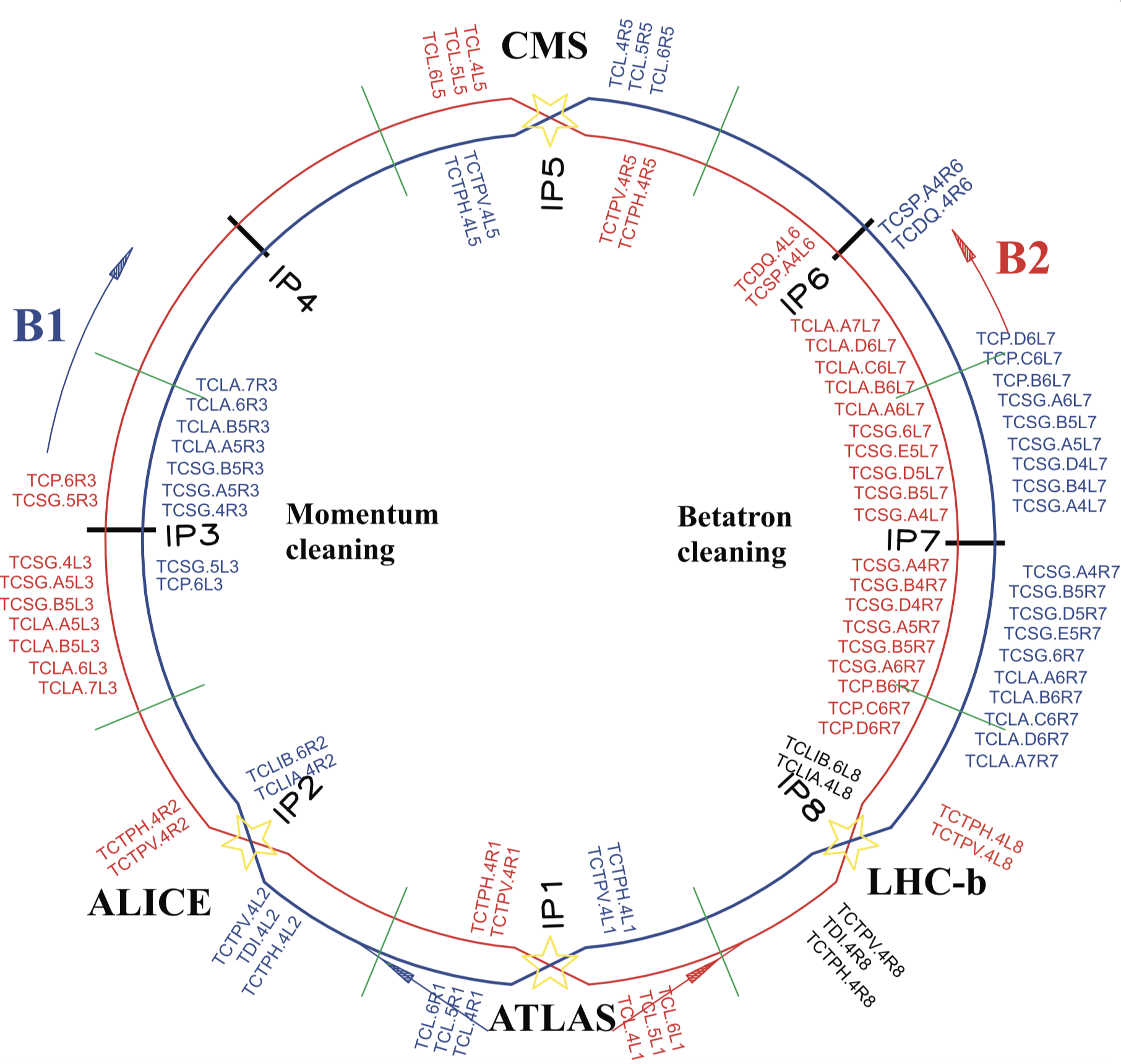}
  \caption{
  Schematics of the LHC Run II collimation layout (prior to 2019)~\citep{bib:w2}.
	}
  \label{fig:1}
\end{figure}

The LHC utilizes a complex multi-stage collimation system, which is mainly located in two designated Insertion Regions (IRs): IR7 for betatron cleaning and IR3 for momentum cleaning~\cite{bib:Ass_1, bib:Ass_2, bib:Bruce}. As shown in Fig.~\ref{fig:1}, it consists of over 100 movable collimators, where most consist of two movable jaws that are aligned symmetrically around the beam~\citep{valentino12}, which is eased by the in-jaw beam position monitors (BPMs) in the most recent design~\citep{valentino17_PRSTAB}. Primary collimators (TCP) constitute the smallest bottleneck of the ring and should intercept large-amplitude halo protons. Secondary collimators (TCSG) should catch the secondary halo leaking out of the TCP. Active absorbers (TCLA) should attenuate showers and intercept part of the tertiary halo. Tertiary collimators (TCTs) in the experimental insertions protect the local aperture bottlenecks and help controlling experimental backgrounds~\cite{bruce13_NIM_backgrounds,bruce19_PRAB_beam-halo_backgrounds_ATLAS}. The primary and secondary collimators are close to the beam  and therefore need to be robust against beam impacts. In the present LHC they are therefore made of carbon-fibre-composite (CFC), which, however, does not have an optimal electric conductivity and therefore gives rise to a high impedance. The tertiary collimators and absorbers are made of a tungsten alloy and usually operated at larger apertures and therefore contribute less to impedance. 


The LHC collimation system follows the transverse size of the beam as it shrinks down during acceleration and the local $\beta$-function change during the optics squeeze before the beams are brought into collision. For that reason at the top energy the collimation system becomes the single highest contributor to the machine transverse impedance \citep{bib:w3}. As the bunch population doubles to $2.3\times 10^{11}$~protons per bunch (ppb) at injection \citep{bib:w4} in the High-Luminosity (HL-LHC) upgrade, the impedance has to be reduced to ensure beam stability for all operational scenarios. In the present paper we focus on the most critical case for single-beam stability, just before the beams are brought into collision, at the beginning of the luminosity levelling process, when the $\beta$-function at the main interaction points reaches $\beta^* = 41$~cm (for the ultimate luminosity of $7.5\times 10^{34}$~cm$^{-2}$s$^{-1}$), and the brightest baseline type of beam.

\begin{figure}[h]
  \centering
  \includegraphics[width = 3.375in]{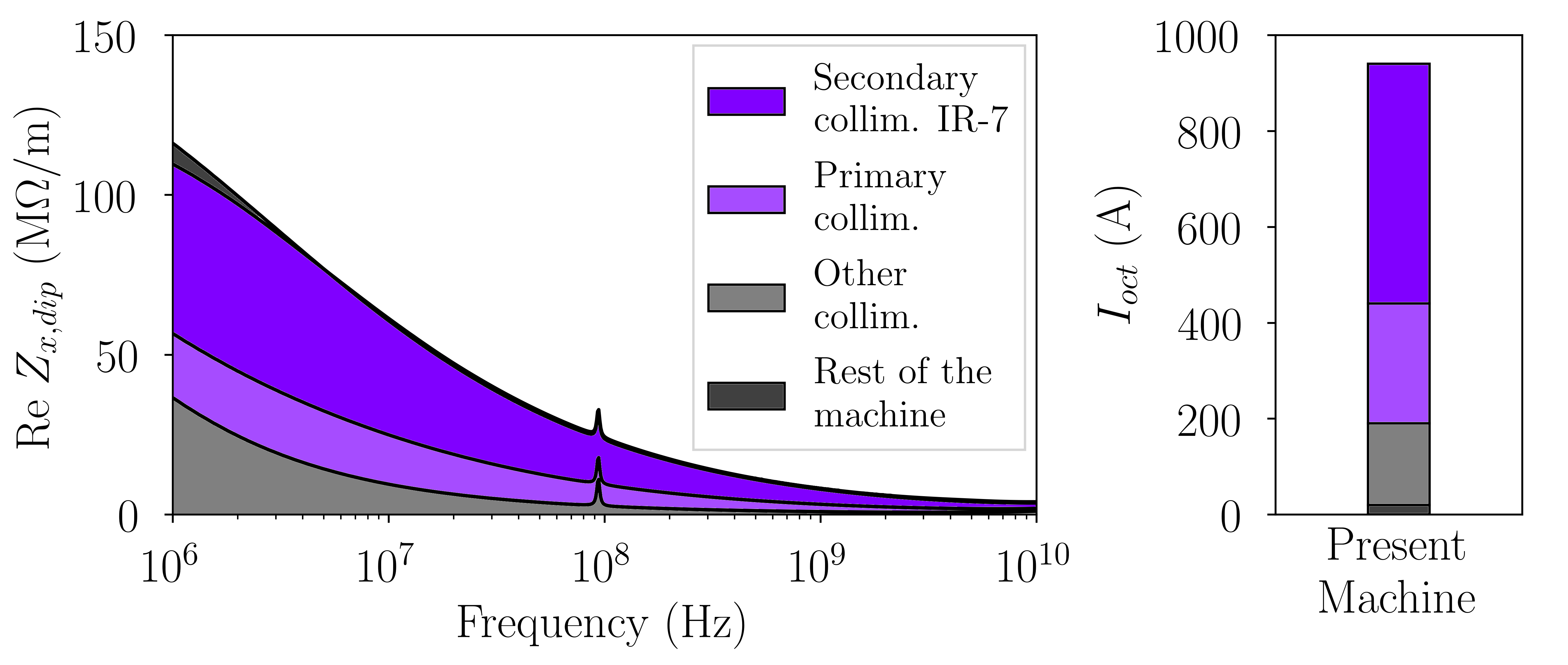}
  \caption{
  LHC collimators dominate the overall machine impedance at the top energy (plotted on the left -- real part of the dipolar impedance as a function of frequency). They are responsible for nearly all the octupole current required to stabilize the beam, with $\sim50$\% coming from 11 secondary betatron cleaning collimators (plotted on the right -- estimated octupole threshold with a factor 2 included, details of the simulation can be found in Sect.~III). Simulation results for $E = 7$~TeV, $Q' = 10$, Bunch Compression, Merging and Splitting (BCMS) beam \citep{bib:BCMS}, Ultimate operational scenario \citep{bib:w9}, the most critical, horizontal plane is shown.
	}
  \label{fig:2}
\end{figure}

The LHC stability is typically quantified in terms of current in its Landau Octupole system, providing Landau damping of collective instabilities. The maximum operating current for which the system has been commissioned is 570~A~\citep{bib:Xavier_Evian}. When estimating the realistic octupole threshold one has to take into account an uncertainty of the impedance model, optics errors \citep{bib:w8}, magnet imperfections, linear coupling \citep{bib:w6}, an uncertainty of the beam distribution and other detrimental effects like the long-range beam-beam interaction \citep{bib:w5} and the transverse feedback noise \citep{bib:w7,Furuseth:2019iym}. Due to these effects in operation the octupole current has to be around a factor two larger than what is predicted for an ideal machine from impedance considerations only~\citep{bib:Xavier_Evian, bib:Elias_Coll_Review}.
For the HL-LHC project this assumption requires a dramatic reduction of the collimator contribution to the octupole threshold (Fig.~\ref{fig:2}). Since the operational collimator openings cannot be significantly relaxed, it is therefore planned to change the material of the collimators with the highest contributions to the impedance. The present upgrade baseline foresees to replace 9 out of 11 secondary and 2 out of 4 primary collimators per beam~\citep{bib:w4} with new collimators using jaws made of a novel low-impedance material~\citep{Guardia-Valenzuela:2018bgf}.

This paper is structured as follows: first, in Sect.~II we introduce the novel low impedance materials chosen for the collimator upgrade and present the results of beam and bench measurements of coating resistivity performed on the prototype low impedance collimator. Thanks to a high resolution attained in the measurements of beam tune shift we demonstrate a reduction of the collimator resistive wall tune shift with each novel material. We compare the improvements with theoretical predictions and discuss the possible sources of discrepancies. Then in Sect.~III we assess the impact of installation of low impedance collimators in the machine and assess coherent stability margins for realistic future operational scenarios using a Vlasov numerical solver. Finally, in Sect.~IV we discuss potential options for further impedance minimization.

\section{Prototype low-impedance collimator}

In order to reduce the transverse impedance of HL-LHC several low resistivity material options have been considered for its collimators. First, the jaws of the most critical primary and secondary collimators can be replaced with molybdenum-graphite (MoGr) that is characterized by a factor of five lower bulk DC resistivity than the presently used carbon fibre composite (CFC): $\rho =$ 1 vs 5~$\mu\Omega$m. On top of that, a jaw can be coated with a thin layer of a low-resistivity molybdenum (Mo) coating with a bulk resistivity of $\rho = 0.053~\mu\Omega$m. A 5~$\mu$m coating thickness is sufficiently greater than the skin depth of the coating at the high frequencies, relevant for the single-bunch dynamics ($\sim 1$~GHz), making the impedance at these frequencies nearly independent of the material behind the coating \citep{bib:th8}.

\begin{figure}[h]
  \centering
  \includegraphics[width = 3.375in]{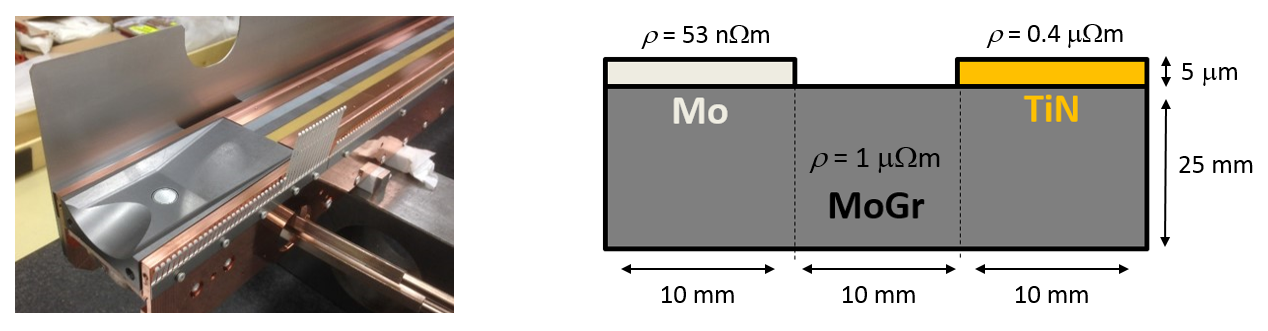}
  \caption{
  The prototype collimator has two 10~mm wide low resistivity (Mo and TiN) stripes on a MoGr substrate.  Left - photo of the collimator assembly; right - schematic drawing of the collimator jaw.
	}
  \label{fig:3}
\end{figure}

To test the novel materials with beam, a special collimator has been installed in LHC. Its design, similar to the baseline design foreseen for the HL-LHC project (Fig.~\ref{fig:3}, left), relies on a modular concept that allows embarking different absorber materials in the jaws, with no other impact or modifications to the other collimator components. This design can thus be adopted indifferently for primary, secondary, and tertiary collimators, which is advantageous for series production. It also features beam position monitors for orbit measurements and alignment. The prototype jaws allow testing three different approaches. The jaws are made of MoGr grade MG-6403Fc~\citep{Guardia-Valenzuela:2018bgf} and one 10 mm wide surface of uncoated material and two 10 mm wide coating stripes of Mo -- the baseline coating of secondary collimators -- and titanium nitride (TiN) with $\rho = 400$~$\mu\Omega$m -- for additional reference measurements  (Fig.~\ref{fig:3}, right). The jaws can move in the non-cleaning transverse plane, exposing the beam to one of the stripes at a time, and thus effectively ``selecting'' the coating material to study. This so-called ``three-stripe collimator'' was installed next to a standard secondary collimator, allowing comparing the performance of its materials with the presently used CFC. The installation slot has been chosen, where the beam size and consequently the collimator gap is smallest thus maximizing the sensitivity of the impedance measurement.

\subsection{Beam Measurements}

\begin{center}
\begin{figure}[b!]
\includegraphics[width = 6.5in]{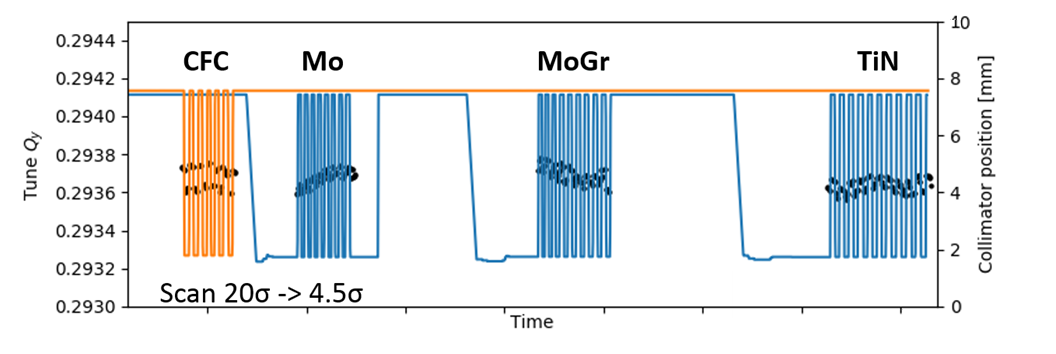}
      \caption{
      The raw tune measurement data shows a clear reduction of the tune shift with the new materials with respect to the CFC. A significant tune drift during the measurement, not related to the collimator movement, can also be seen. The orange line depicts the position of the jaws (full gap) of the standard CFC secondary collimator, the blue line -- the prototype collimator. Black dots show individual tune measurements.
      }
  \label{fig:4}
\end{figure}
\end{center}

A relevant measure that quantifies each material is the magnitude of the resistive wall tune shift, created when the collimator jaws are brought closer to the beam. To measure the tune shift the collimator gap was cycled between a large gap, where the collimator impedance is negligible, and a small gap of 4-6 reference beam sizes. At each gap transverse beam oscillations were excited by the transverse feedback system \citep{bib:Hofle_HLLHC_15} (Fig.~\ref{fig:4}). Two separate measurements were performed with single bunches of nominal, i.e. $1.2\times 10^{11}$~p, and high intensity, $1.9\times 10^{11}$~p, at 6.5 TeV (Table \ref{tab:1}). In both tests chromaticity and octupole current were optimized to increase the decoherence time to about 1000 revolutions, which allowed accurately determining the tune at each collimator opening with the SUSSIX \citep{bib:th9} algorithm, while ensuring the transverse stability of the circulating bunch.

Typically, a standard CFC secondary collimator creates a tune shift up to $\sim 10^{-4}$ for a $\sim 10^{11}$~p bunch and collimator openings of interest in LHC. The three low-impedance materials are expected to produce tune shifts two to ten times lower. In order to resolve such a tune shift, one has to be able to measure the tunes with a precision level of $10^{-5}$. One of the challenges is the drift of the tune over the period of the measurement, arising from temperature fluctuations or the noise in the orbit feedback system. In LHC the magnitude of this slow tune jitter with a timescale of the order of 100~s, can be as large as $10^{-4}$ \citep{bib:th10}, which is significantly greater than the expected tune shift of the best coatings. The tune drift can be removed from the data using a special measurement procedure where the collimator gaps were cycled rapidly between their open and closed positions while continuously exciting the beam and measuring its tune (App.~\ref{sec:App_procedure}).


To separate the resistive wall component of the tune shift from the geometric one, an input from a numerical LHC impedance model is used \citep{bib:w3,bib:Nicolo_IPAC}. The model treats the geometry of collimator transitions in the flat taper approximation \citep{bib:th11}, which was found to be in good, 10--15\% level agreement with numerical simulations (see App.~\ref{sec:Geom_tune_shift}). Under the flat-taper approximation the geometric tune shift $\Delta Q_y^{geom}$ is inversely proportional to the square of the gap $\Delta Q_y^{geom} \propto 1/g^2$ \citep{bib:th12}. In the frequency regime of interest the resistive wall component has a steeper dependence on the gap:
\begin{equation} \label{eq:2}
	\Delta Q_y^{RW} \propto \sqrt{\rho}/g^3,
\end{equation}
where $\rho$ is the electrical DC resistivity of the jaw material.

\onecolumngrid
\twocolumngrid

\begin{table}
  \centering
  \begin{center}
  \caption{Key parameters of the measurement with nominal intensity LHC beam (in parenthesis -- for a high-intensity beam). In collimator settings, $\sigma$ is the beam size for a 3.5~$\mu$m reference normalized emittance.}
  \label{tab:1}
    \begin{tabular}{lr}
      \hline\hline
      Parameter & Value \\
      \hline 
Beam energy & 6.5 TeV \\
Bunch intensity & $1.2~(1.9)\times 10^{11}$~p\\
Normalized emittance & $2.0~\mu$m, rms\\
Bunch length & 8.1~cm, rms\\
Chromaticity, h \& v & 7, 7\\
Octupole current & 270 A\\
Coll. retraction cycle & $20\sigma$ to $3.5 - 6.0\sigma$\\
\hline 
\hline 
\end{tabular}
\end{center}
\end{table}

\begin{figure}
  \centering
  \includegraphics[width = 3.375in]{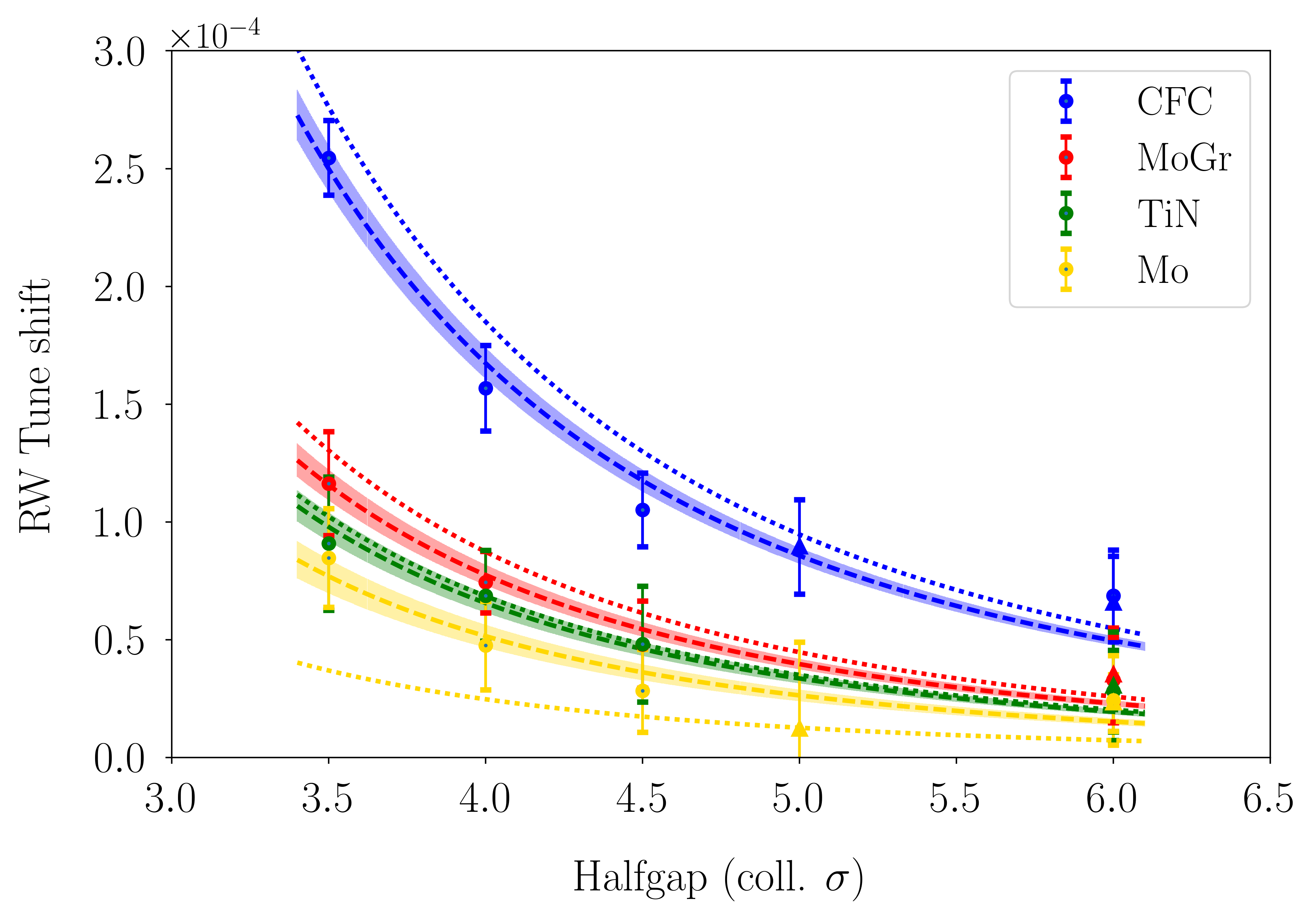}
  \caption{
  The use of uncoated MoGr (red) reduces the resistive wall tune shift compared to the uncoated CFC (blue); each type of coating: TiN (green) and Mo (yellow) further improves the conductivity and can be clearly differentiated. For most materials the results are within 10-20\% of the model predictions (dotted lines). Dots and triangles show the measured data obtained for the nominal intensity of $1.2 \times 10^{11}$~p and the high intensity of $1.9 \times 10^{11}$~p (scaled down to the nominal intensity) respectively. The dashed curves represent Eq.~(\ref{eq:2}) fits with their $\pm 1$~rms uncertainties.}
  \label{fig:6}
\end{figure}

Accounting for the geometric tune shift and fitting the data with Eq.~(\ref{eq:2}) one can clearly distinguish between the different coating options and assess their benefits (Fig.~\ref{fig:6}). A significant decrease of the resistive wall tune shift compared to CFC is observed for MoGr and each type of coating. The largest reduction, as expected, is measured for the Mo coating that has the lowest resistivity. In order to compare with theoretical predictions the expected tune shifts have been computed using the IW2D software under approximation of parallel plate geometry. The IW2D code is based on field matching techniques and computes driving and detuning impedances for an arbitrary number of layers with different material properties \citep{bib:th11}.

The fitted experimental data for CFC, MoGr bulk, and TiN agree with the predictions of the LHC impedance model within 10 to 20\%. A larger discrepancy, up to a factor of two is observed for the Mo coating, indicating a possibly larger than expected resistivity of the coating. Table~\ref{tab:2} summarizes the findings in terms of effective resistivity.

\subsection{Investigating the Higher Resistivity of the Mo Coating}

\begin{figure}
  \centering
  \includegraphics[width = 1.5in]{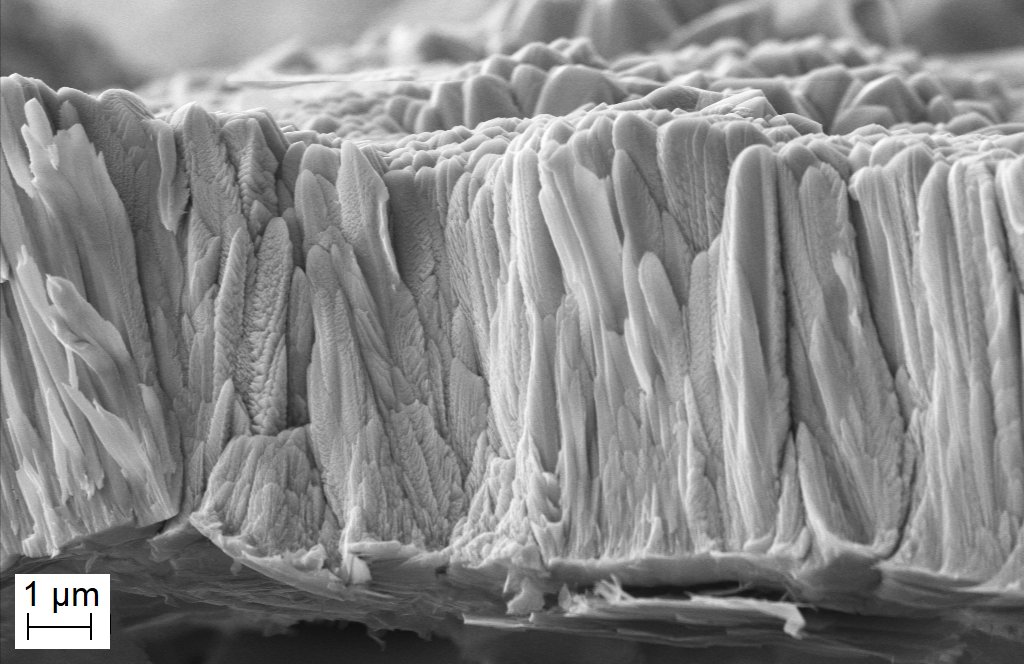}
  \includegraphics[width = 1.5in]{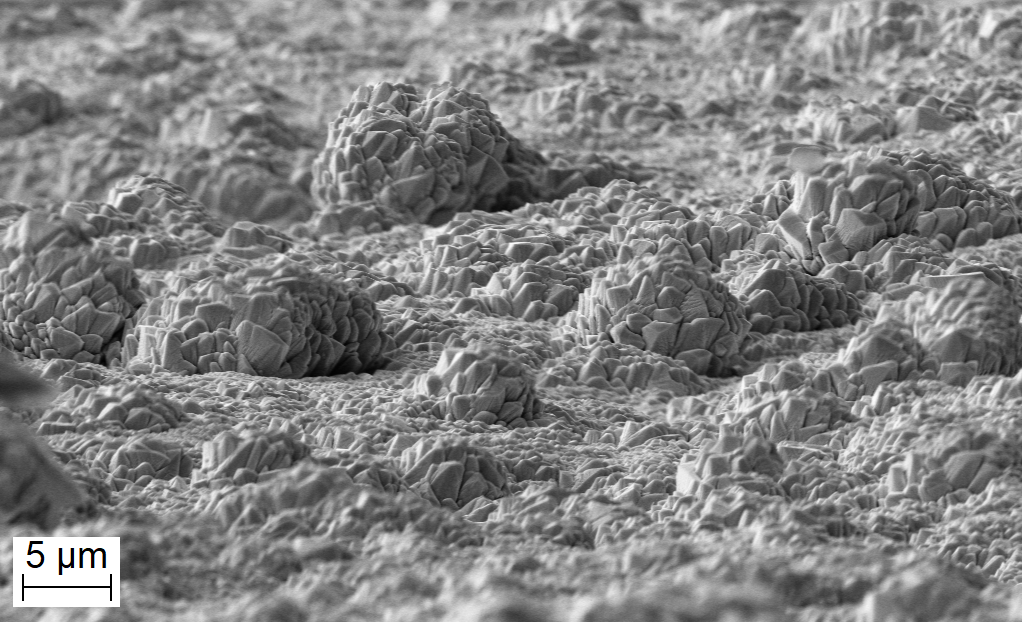}
  \caption{
  SEM imaging \citep{bib:th15} reveals that Mo coating is not uniform: it has a column-like fine structure (left) and inhomogeneities up to 10~$\mu$m on its surface (right) that may affect the measured tune shift.}
  \label{fig:7}
\end{figure}

Several physical effects may contribute to the higher than expected tune shift observed in the Mo-coating. First, the Mo coating has a column-like microstructure with the grains having in-plane sizes below 0.5~$\mu$m (Fig.~\ref{fig:7}, left); the size of the columns decreases for thinner films, increasing the number of transitions an electron crosses when moving in the material and thus increasing the resistivity. Four-point measurements show a significant increase of Mo thin film resistivity at or below the thickness of 5~$\mu$m \citep{bib:th13}; high DC resistivities have been measured in some Mo-coated samples at CERN \citep{bib:Adnan}.

SEM imaging also shows significant roughness of the coated surface: the average size of inhomogeneities is of the order of several microns and is measured to be up to 10~$\mu$m for the test sample with 8~$\mu$m coating thickness (Fig.~\ref{fig:7}, right). Such roughness, not seen in other coatings, should lead to an increase of the imaginary part of impedance in the long-bunch limit \citep{bib:th14}. The additional imaginary impedance scales as $\sim 1/g^3$ making it thus indistinguishable for the resistive wall component in the measurement~[see Eq.~(\ref{eq:2})]. The effect though is rather small -- at least an order of magnitude lower than the expected restive wall impedance even for a large size of roughness ``bumps'' of 5~$\mu$m, similar to the thickness of the coating.

\begin{figure}
  \centering
  \includegraphics[width = 2.5in]{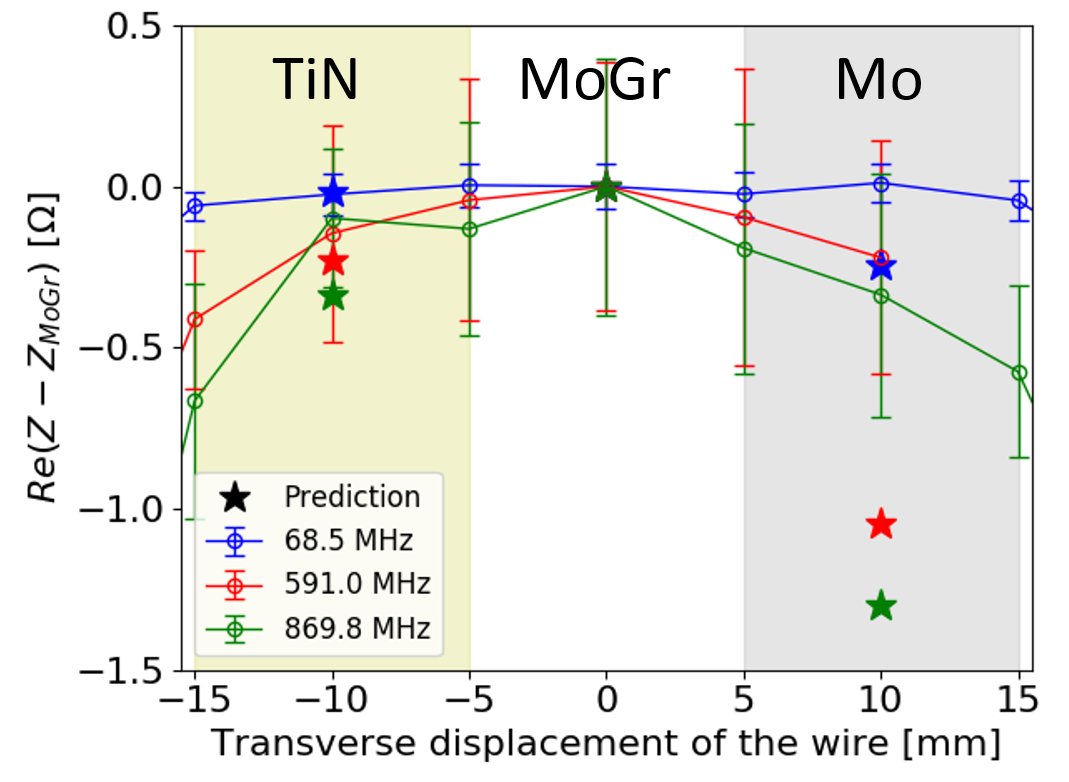}
  \caption{
  A resonant wire measurement of the jaw impedance was performed at different locations along the jaw in a test stand prior to the installation of the prototype in LHC~\citep{Biancacci:2018cxa}. The difference in the real part of the longitudinal impedance with the respect to the uncoated bulk, depicted by lines and error bars, suggests a larger than expected resistivity of the Mo stripe for all tested frequencies: 68.5~MHz (blue), 591.0~MHz (red), and 869.8~MHz (green). The expected values are based on numerical simulations with IW2D \citep{bib:th11} and shown by stars.}
  \label{fig:8}
\end{figure}

The hypothesis of the influence of the microstructure was supported by a RF resonant wire measurement, performed on the three-stripe collimator on a bench at several frequencies relevant for single bunch dynamics. In this test a wire has been horizontally shifted on top of each stripe and the corresponding real component of the impedance has been computed from the change in the quality factor \citep{Biancacci:2018cxa}. This method measured a slightly greater impedance than the one simulated with IW2D, although the discrepancy was likely caused by a constant additional resistive wall impedance of the tapered transitions, which are made of MoGr. The variation of the real part of the longitudinal impedance with respect to the bulk MoGr matched the expected values within uncertainties for the TiN stripe, while the Mo stripe showed a lower than expected impedance difference with respect to the bulk (Fig.~\ref{fig:8}). This result suggests an extra resistivity of the Mo coating, which is consistent with the results of beam measurements. Similar results have also been obtained by resistivity measurements performed at DC on samples of coated and uncoated material using a 4-wire technique \citep{bib:Adnan}. Table~\ref{tab:2} provides a summary of all the measurements done and the material resistivities that can be assumed from their results. 

The prototype coating was created using a standard magnetron sputtering process. Further investigations, including DC and RF measurements for various substrates and different coating procedures are under way \citep{bib:Adnan}. Preliminary results of those studies indicate an improvement of coating resistivity to below $0.07~\mu\Omega$m, close to that of pure Mo, which is achieved with good reproducibility when using a high-power impulse magnetron sputtering process \citep{bib:Adnan}. This method is now being used for the series production. Nevertheless, a potential larger coating resistivity up to $0.25~\mu\Omega$m (which corresponds to what has been measured with beam in LHC) is also taken into consideration for stability analysis. 

\begin{table}
  \centering
  \begin{center}
  \caption{Comparison of the measured and expected resistivities (n$\Omega$m). In the RF test, the materials are measured relative to MoGr, which is assumed to have the nominal resistivity.}
  \label{tab:2}
    \begin{tabular}{lccc}
      \hline\hline
      Material & Model & Beam & Lab: RF\\
      \hline 
CFC & 5000 & $4030 \pm 380$ & --\\
MoGr & 1000 & $760 \pm 60$ & (1000)\\
TiN & 400 & $340 \pm 40$ & 400\\
Mo & 53.4 & $250 \pm 50$ & 300\\
\hline 
\hline 
\end{tabular}
\end{center}
\end{table}

\section{OUTLOOK FOR HL-LHC}

\begin{figure}
  \centering
  \includegraphics[width = 2in]{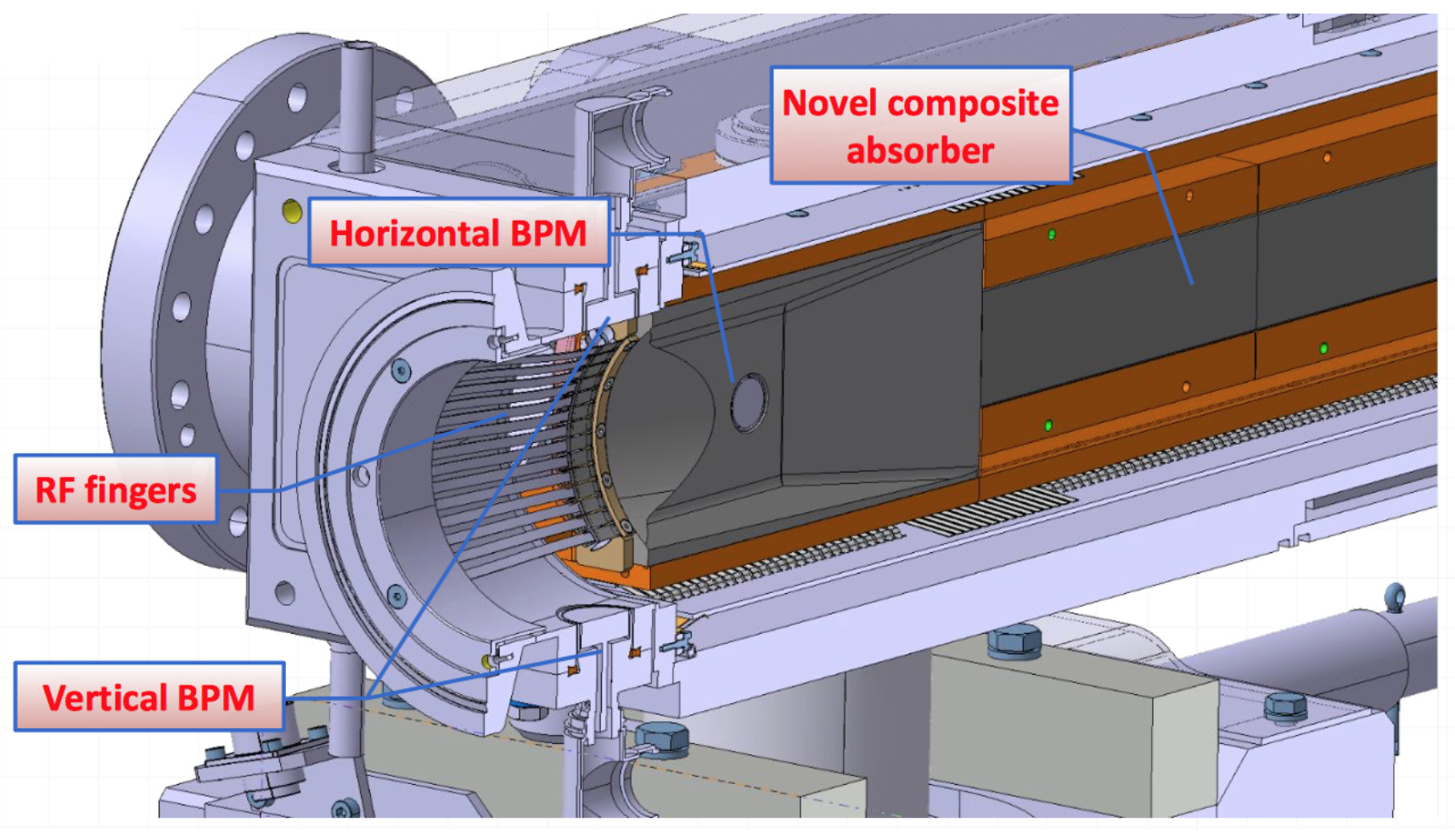}
  \caption{
  3D jaw layout of the novel secondary collimator design \cite{fedeColUSM}.}
  \label{fig:tcspm}
\end{figure}

The present baseline of the HL-LHC collimator impedance upgrade foresees that a total of 9 out of 11 secondary betatron cleaning collimators per beam will be replaced. The new design follows that of the three-stripe prototype: a MoGr active part coated with 5-6 $\mu$m Mo layer. It also includes two in-jaw BPMs for collimator alignment and a BPM for orbit measurements in the plane orthogonal to the collimation plane (Fig.~\ref{fig:tcspm}). Details of other design improvements can be found in \cite{fedeColUSM}. In addition to the secondary collimator upgrade, four betatron primary collimators (1 per beam per plane) will be replaced with the uncoated MoGr ones. 

\subsection{Impact on transverse beam stability}

The effect of low-impedance collimators on the transverse beam stability has been estimated using the HL-LHC impedance model \citep{bib:Nicolo_IPAC} and the latest beam and optics parameters (Table~\ref{tab:3}). The simulations were performed with Vlasov solvers NHT \citep{bib:NHT} and DELPHI \citep{Mounet:2018wni,bib:w13}, capable of treating combined head-tail and coupled-bunch motion. The codes determine the coherent tune shift of the most unstable mode, which is then converted into the octupole strength required to stabilize that mode using a stability diagram approach and assuming the modes are independent (far from the transverse mode coupling instability (TMCI) threshold).

To find the octupole threshold we, first, compute the nonlinear detuning, required to stabilize impedance-driven instabilities using a stability diagram approach. The diagrams are calculated for a pessimistic case, where the tails of the transverse distribution are cut at $3.2\sigma_{rms}$ \citep{bib:w14}, and assuming no emittance blow-up at injection (Table~\ref{tab:3}). The octupole thresholds are then computed from the detuning, neglecting the enhancement of the tune footprint due to telescopic optics \citep{bib:w15} or second order chromaticity \citep{bib:Sec_Order_Chroma} and the detrimental long-range beam-beam interaction \citep{bib:w5,bib:BBNote}. The most critical baseline beam type is examined: bunch compression, merging and splitting (BCMS), which is prepared using a special procedure in the injectors and has slightly smaller number of bunches and transverse emittance than the standard beam \citep{bib:BCMS,bib:w9}.

\begin{figure}
  \centering
  \includegraphics[width = 3.25in]{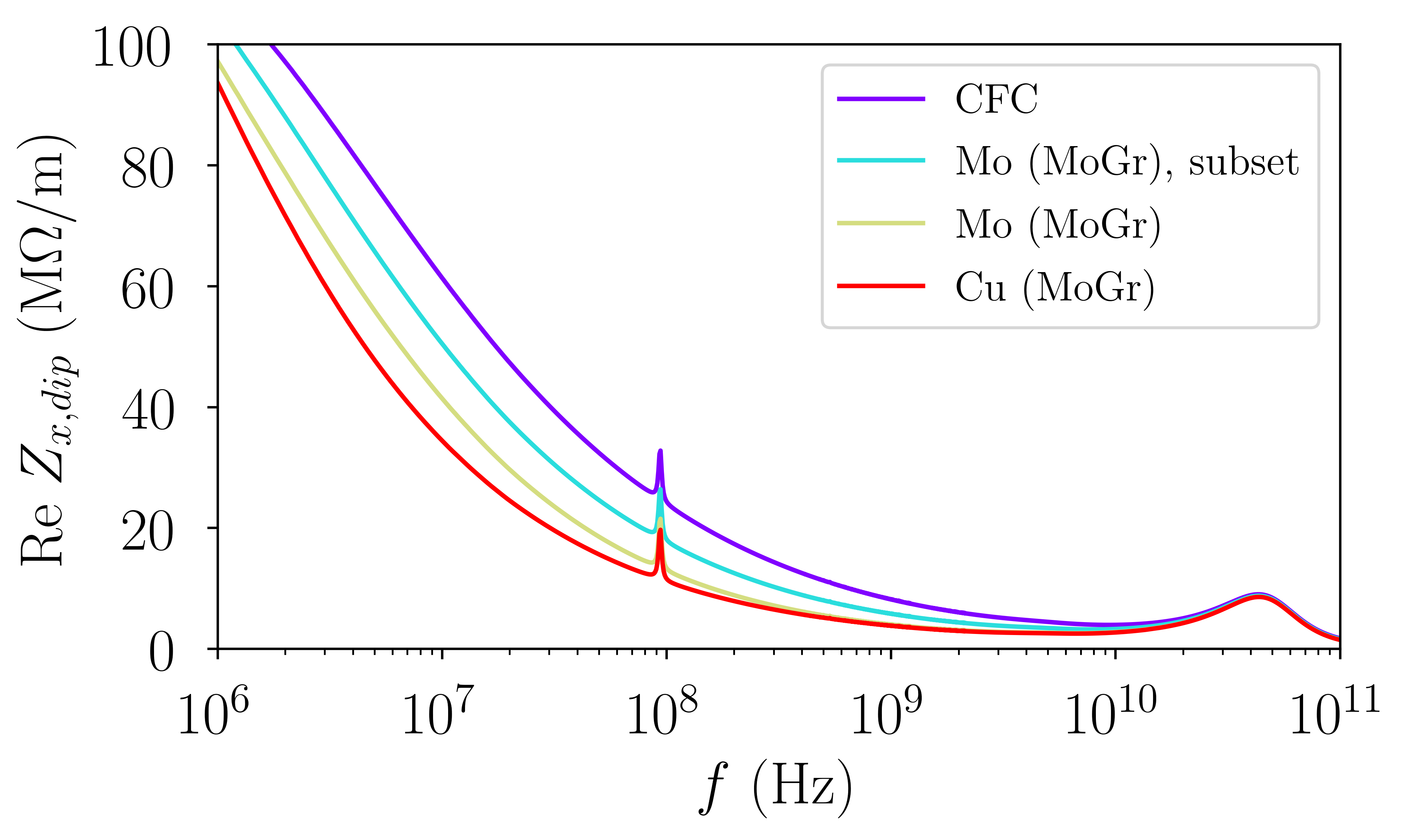}
  \caption{
  Low-impedance secondary collimators decrease the machine's horizontal dipolar impedance (real part plotted along the vertical axis) at the top energy by 30\% at the frequencies around 1~GHz, relevant for the single-bunch coherent beam dynamics. Coating a subset of four collimators provides a half of the reduction. Energy 7~TeV; higher order modes (HOMs) of HL-LHC crab cavities not shown for simplicity.}
  \label{fig:9}
\end{figure}

The greatest impact on beam stability is expected from the coating of the secondary collimators due to their large share of the octupole threshold. The reduction of machine impedance due to upgrading the IR-7 secondaries alone is shown in Fig.~\ref{fig:9}. Since low-frequency coupled-bunch instabilities can be efficiently suppressed by the transverse feedback, the threshold is governed by the high frequency part of beam impedance, relevant for head-tail instabilities, above the RF frequency of 400 MHz. Upgrading the collimators reduces it by 30\%, and a half of the total impedance reduction is obtained by coating a subset of four collimators \citep{bib:w18}, chosen for LS2 (Fig.~\ref{fig:9}). 

\begin{table}
  \centering
  \begin{center}
  \caption{Key beam and machine parameters used for numerical simulations. Collimator settings are defined for a $2.5~\mu$m rms reference normalized emittance.}
  \label{tab:3}
    \begin{tabular}{lr}
      \hline\hline
Beam energy, $\beta^*$ & 7~TeV, 41~cm \\
Number of bunches & 2748\\
Bunch intensity & $2.3\times 10^{11}$~ppb\\
Tunes: x, y, z & 62.31, 60.32, $2.1\times 10^{-3}$\\
Normalized emittance & $1.7~\mu$m, rms\\
Bunch length & 9.0~cm, rms\\
Feedback damping time & 100 turns\\
Chromaticity & 10\\
Primary coll. settings & $6.7\sigma$\\
Secondary coll. settings & $9.1\sigma$\\
\hline 
\hline 
\end{tabular}
\end{center}
\end{table}

\subsection{Emulation in LHC}

\begin{figure}
  \centering
  \includegraphics[width = 3.25in]{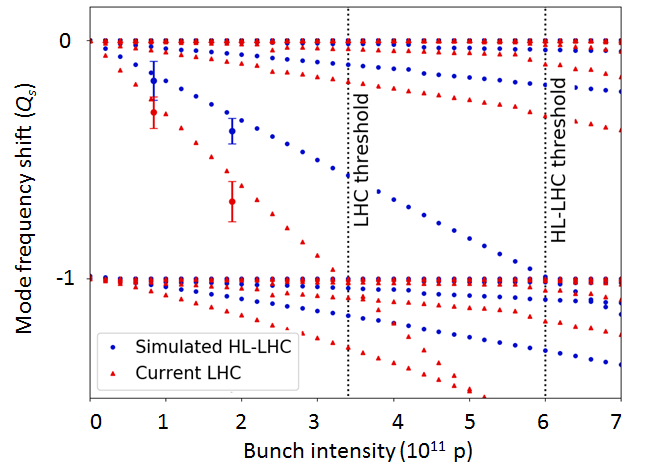}
  \caption{
  The collimator upgrade is expected to increase the TMCI threshold at the top energy by nearly a factor of two in HL-LHC (blue) compared to LHC (red). The measured mode frequency shifts (error bars) are in good agreement with simulation predictions (dotted lines). Beam~1, $E = 6.5$~TeV, $Q' = 5$ \citep{bib:w17}.}
  \label{fig:TMCI}
\end{figure}

In LHC, the beam intensity is predicted to be limited around $3.4\times 10^{11}$~ppb by the coupling of modes 0 and -1 in the horizontal plane, which for zero chromaticity and in the absence of the transverse feedback causes the fast TMCI instability \citep{bib:w17}. The present threshold estimate is in good agreement with the measurements of mode 0 tune shift with bunch intensity (Fig.~\ref{fig:TMCI}). The deployment of low-impedance secondary collimators will increase the threshold to about $6.0\times 10^{11}$~ppb for the same collimation settings, nearly doubling the threshold and providing enough margin for the HL-LHC high intensity beam. The low-impedance collimators were emulated by a corresponding increase of the gap of the existing ones. According to Eq.~(\ref{eq:2}) the gaps were relaxed by a factor 2.1, from 6.5 collimation $\sigma$ used normally to 14 $\sigma$ (defined for $3.5~\mu$m reference normalized emittance for LHC). A measurement of mode 0 tune shift is again in good agreement with the impedance model predictions, confirming a significant reduction of the machine impedance (Fig.~\ref{fig:TMCI}).

\subsection{Staged Collimator Upgrade}

The HL-LHC project strategy is to pursue a staged deployment of the low-impedance collimators, consisting of two phases: a first installation in the Long Shutdown 2 (LS2, in the period 2019--2020) \citep{bib:w4} followed by a second installation in LS3, in 2024--2026. This approach has various advantages as explained in detail in Ref.~\cite{bib:w18}. It already provides an important reduction of the collimator impedance for the LHC Run~3, when brighter beams progressively become available thanks to the LHC Injector Upgrade (LIU) program~\citep{bib:LIU}. This will provide important benefits to the LHC operation and will allow studying better the possible impedance limitations. In addition, a staged deployment allows possible further iterations on the new collimator design for the second production line for LS3. 

For an optimum deployment of low-impedance collimators in Run~3, various studies were carried out to identify the IR7 secondary collimator slots to be upgraded with highest priority. This analysis started with an assessment of the slots that contribute most to the collimator impedance and also included the overall performance of the collimation system, the beam loads and the subsequent thermo-mechanical responses of the jaws that may affect beam lifetime, and the exposure of the hardware to failure scenarios at the injection and at the top energy.
A solution excluding the replacement of the collimators that are the most loaded in case of regular collimation losses (in terms of energy deposition) has been chosen. This option also features the largest impedance reduction in the most critical horizontal plane \citep{bib:w18}.  

\begin{figure}
  \centering
  \includegraphics[width = 3.25in]{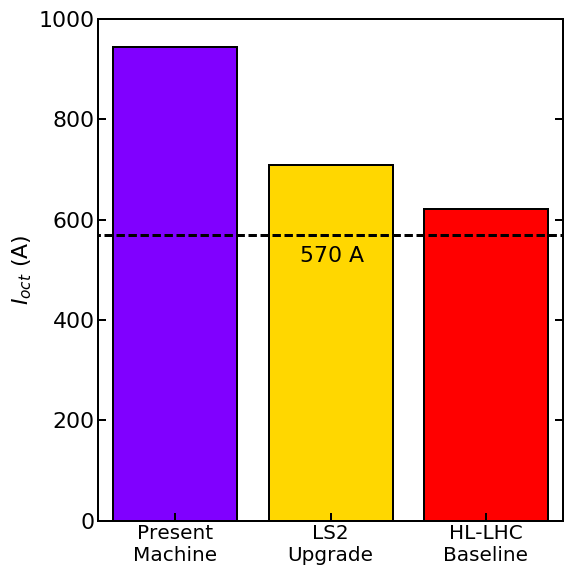}
  \caption{
  Novel coatings significantly improve the single beam octupole threshold. For the most critical BCMS beam up to $\sim 320$~A is gained by upgrading 9 secondary and 2 primary collimators per beam in the present baseline (``HL-LHC Baseline'') compared to the current machine (``Present Machine''). ``LS2 Upgrade'' reflects the situation in Run~3 with 2 primary and 4 secondary collimators upgraded; it provides over a half of the overall improvement, about 250~A. $E = 7$~TeV, $Q' = 10$, the situation in the most critical, horizontal plane is shown, assuming a factor 2 from the operational experience.
  }
  \label{fig:10}
\end{figure}

Analyzing potential options one can see that, first, the complete upgrade foreseen by the HL-LHC project significantly lowers the octupole threshold: by over 300~A, or about 1/3 of the current. It brings the octupole current close to the operational limit for the most challenging operational scenario (Fig.~\ref{fig:10}). Second, the chosen first-stage upgrade option (2 primary and 4 out of 11 secondary betatron cleaning collimators per beam) provides more than a half of the overall octupole current reduction: nearly 250~A (Fig.~\ref{fig:10}). The improvements become somewhat smaller if one assumes the Mo resistivity from the beam measurements. For the LS2 upgrade the current increases by 30~A, or around 4\%, and for the HL-LHC baseline -- by 50~A, or less than 10\%. 


\section{Ways to further reduce the impedance}

As the resistive wall part of the impedance is reduced thanks to the low-impedance coatings, it now becomes important to model more accurately other sources of impedance, in particular the geometric impedance of the collimators. For the full collimator upgrade the total collimator resistive wall component amounts to 8.0~M$\Omega$/m (54\%) and the total geometric -- to 3.5~M$\Omega$/m (24\%) out of 14.8~M$\Omega$/m overall effective machine dipolar impedance in the vertical plane and 7.6~M$\Omega$/m (46\%) and 5.5~M$\Omega$/m (33\%) out of 16.5~M$\Omega$/m in the horizontal plane, respectively. The remaining 20\% come from various sources, predominantly the beam screens: their resistive wall impedance and the broadband impedance of the pumping holes. In the following paragraphs we provide a brief overview of potential methods to further reduce collimator impedance.

\subsection{Momentum cleaning collimators}

\begin{figure}
  \centering
  \includegraphics[width = 3.4in]{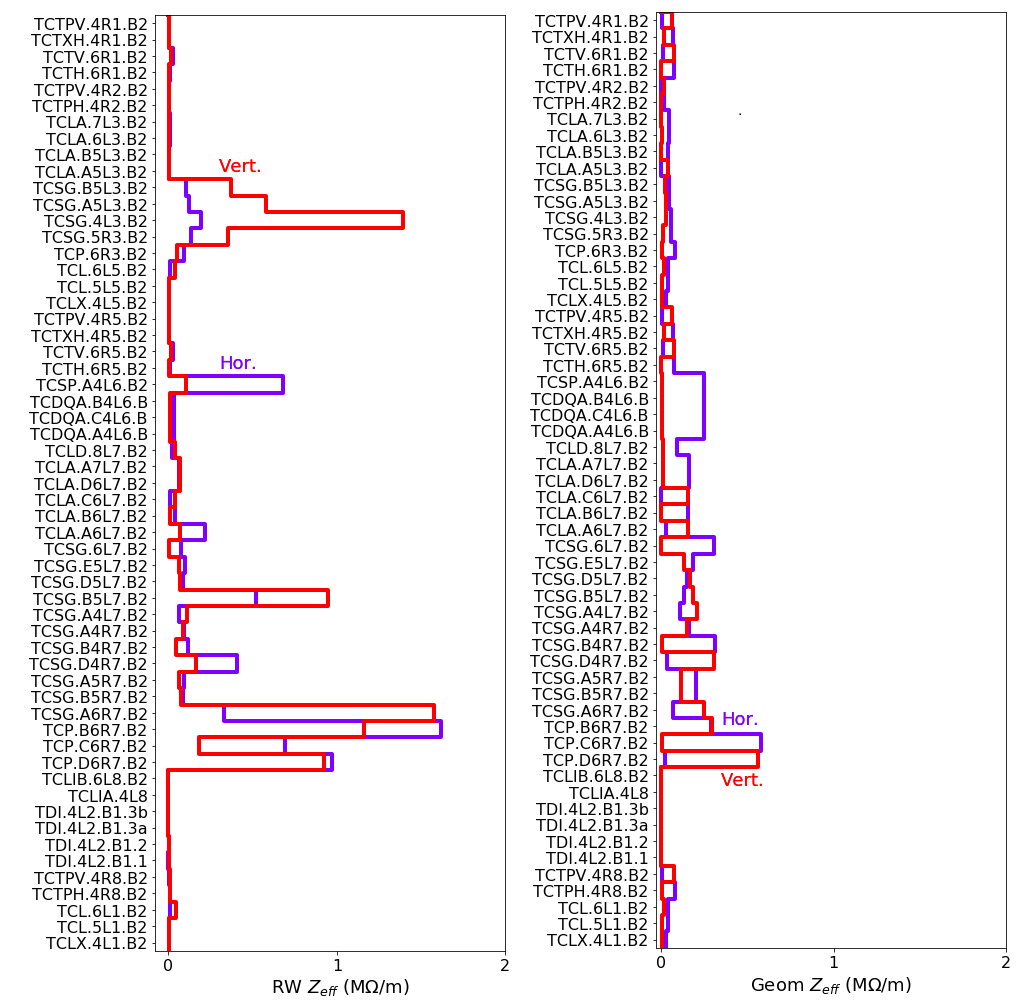}
  \caption{
  Breakdown of the overall dipolar effective imaginary impedance of the machine after the low impedance collimator upgrade by the individual collimators: resistive wall contributions are shown on the left and geometric -- on the right. The upgrade includes 2 primary and 9 secondary collimators per beam. Simulations in IW2D \citep{bib:th11}, top energy $E = 7$~TeV, $\beta^* = 41$~cm, $Q' = 10$, Beam 2. Beam 1 collimators have similar impedance. Primary collimator names begin with `TCP', secondary -- `TCSG', tertiary -- `TCT'.} 
  \label{fig:11}
\end{figure}

Figure~\ref{fig:11} depicts individual collimator contributions to the RW (left) and geometric (right) parts of effective imaginary dipolar impedance at flat-top. RW contributions are computed assuming the current baseline scenario \citep{bib:w9}. Most of the RW contribution comes from three sources: the primary collimators, the secondary collimators in IR-7 and in IR-3. The momentum cleaning secondaries in IR-3 show extra potential for impedance reduction, since they are not upgraded in the baseline, but could be replaced with low-impedance collimators if needed. The upgrade of IR-3 secondaries would further reduce the machine impedance, mainly in the vertical plane, by $\sim 3.5$~M$\Omega$/m.

\subsection{Cu coating}

Copper, having a factor 3 larger DC conductivity than molybdenum, can further significantly reduce the resistive wall component of an individual collimator. But since the overall impedance of the machine is also affected by many other sources, such as the resistive wall impedance of its beam screens or the geometric collimator impedance, Cu coating of the collimators only marginally decreases the overall impedance of LHC (Fig.~\ref{fig:9}). The downside of the coating is its lower tolerance to beam losses compared to Mo, which was observed in HiRadMat tests at CERN \citep{bib:w10}. Nevertheless, the coating might still be used in certain collimators based on the outcome of energy deposition and failure scenarios studies.

\subsection{Optimal taper geometry}

\begin{figure}
  \centering
  \includegraphics[width = 3.5in]{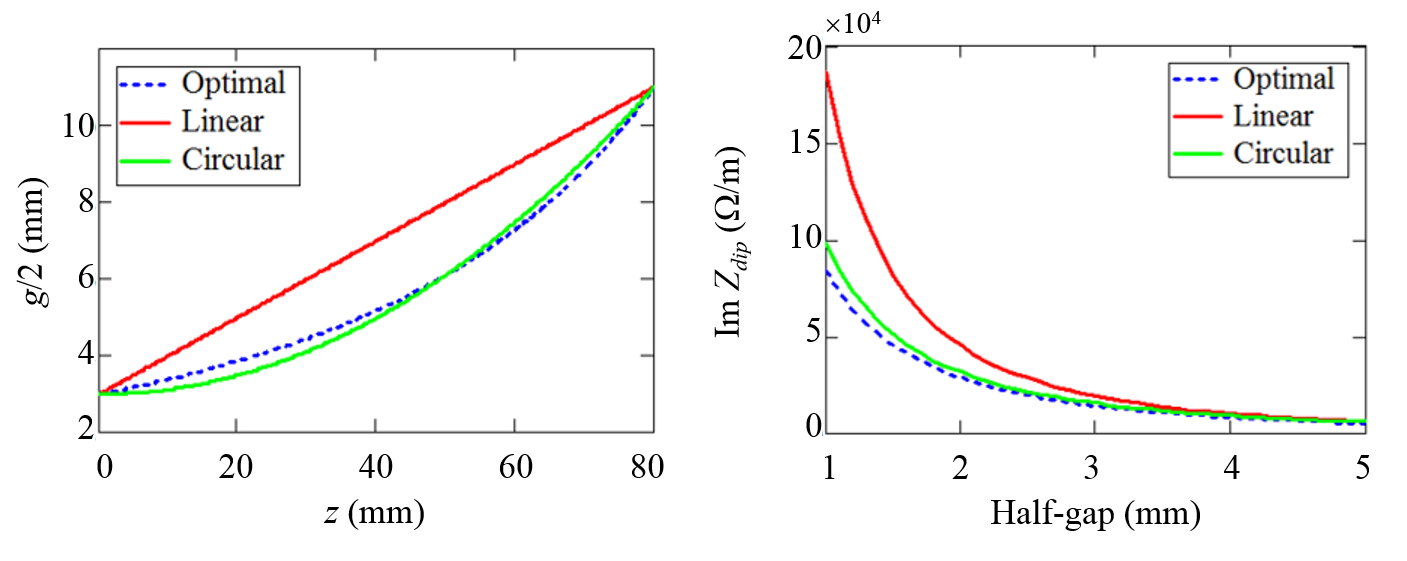}
  \caption{
  A simple round transition follows closely the optimal shape (left) and provides a near-optimal reduction of the geometric broadband impedance in the whole range of collimator openings of interest (right). 5.71~deg transition of the LHC secondary TCSPM collimators with $L = 80$~mm, $w = 80$~mm.}
  \label{fig:13}
\end{figure}

Taper transitions of LHC collimators have already been optimized in order to lower their geometric impedance \citep{Frasciello:2014zda}. The new double taper design of the transitions with a smaller tapering angle closer to the beam offers a factor two decrease of the broadband imaginary impedance of the tapers (see App.~\ref{sec:Geom_tune_shift}). 
A further gain can be achieved by using an optimal non-linear geometry as suggested by \citep{bib:podob}. The shape is designed such as to minimize the geometric impedance contribution of a taper profile $g(z)$:
\begin{equation} \label{eq:geom_imp}
    \centering
        Z_{dip} = -\frac{i \pi w}{c} \int_{0}^{L}\frac{1}{g^3}\left(\frac{\partial g}{\partial z}\right)^2 dz
\end{equation}
for a given tapering length $L$ and width $w$ \citep{bib:th12}. The resulting profile follows 
\begin{equation} \label{eq:optimal_profile}
    \centering
        g(z) = g_0\times \left[ 1 - \frac{z}{L} \left( 1 - \sqrt{\frac{g_0}{g_0 + \Delta g}} \right) \right]^{-2},
\end{equation}
where height $\Delta g = g(0) - g(L)$ stands for the height of the transition and $g_0 = g(0)$ -- the collimator half-gap. Simple estimates show that this approach can further lower the geometric impedance by up to a factor two depending on the gap (Fig.~\ref{fig:13}). The downsides of this approach might be that the shape remains optimal only for one specific collimator opening and that it is rather complex, i.e. may be costly to manufacture.

A simpler similar more viable shape could be obtained for example with an arc of a circle. Considering, for example, the 5.71~deg linear transition of the secondary TCSPM tapers that is the closest to the beam with $L = 80$~mm, $\Delta g = 8$~mm, one can see that its optimal shape can be approximated with arc of a circle of a $R = 80$~mm radius. The arc provides a comparable impedance reduction in a wide range of practical collimator openings (Fig.~\ref{fig:13}). The improvement can be as large as a factor two for sufficiently small collimator openings, or up to $0.2-0.3$~M$\Omega$/m, corresponding up to $\sim 5$~A of octupole current for the BCMS beam. The overall reduction of the octupole threshold depends on the exact number and locations of collimators to be upgraded and goes beyond the scope of this article.

\subsection{Additional collimator retraction}

Since the resistive wall impedance of the collimators is a steep function of their gap, $\propto 1/g^3$, an intuitive way to lower it is by retracting the collimators. This has only a limited impact on the overall machine impedance though due to collimator impedance being already relatively low after the low-impedance upgrade and the impact of other impedance sources, i.e. beam screen. On top of that, this process has significant associated risks limiting the reach in $\beta$-function at the integration points ($\beta^*$). For the LHC a smaller $\beta^*$ results in larger $\beta$-functions in the final focusing system, increasing the risk of unacceptable losses if the aperture is not shadowed by the collimation system. Therefore, to ensure protection, the collimation system must be sufficiently tight, as explained in Refs.~\citep{Bruce:2015qka, Bruce:2017ezf}. Quantitatively, a retraction of the collimation hierarchy by an additional $1~\sigma$ (corresponding to $2.5~\mu$m normalized emittance) could potentially yield up to $\sim 40$~A reduction of the octupole threshold for the BCMS beam with a greater improvement for the previously discussed partial secondary collimator upgrade. But the implications on machine protection have to be carefully studied to ensure adequate protection of the triplet aperture in the most critical scenarios and maintaining an acceptable level of beam losses seen by the equipment. Such a study is currently ongoing.


\subsection{Alternative optics}
Since the impedance of the collimators depends on their opening and thus on the physical beam size, it may be possible to optimize the optics in the collimation regions in order to reduce their impedance. For example, resistive wall impedance of a collimator jaw in the plane of collimation is proportional to the $\beta$-function at its location and inversely proportional the a cube of the gap, which, in turn, is proportional to the beam size. Therefore the impedance scales as
$\beta / \beta^{3/2} = 1 / \sqrt{\beta}$. A study aimed to optimize IR7 optics is currently under way for LHC Run 3 operation~\citep{bib:Run_Alt_Optics}. First results suggest that an improvement around 20\% in terms of overall machine impedance can be achieved while also improving the cleaning efficiency. Experimental validation of the new optics is planned after the restart of LHC in Run~3.

It has to be noted as well that $\beta$-functions at the Landau octupoles can be boosted via the ATS optics~\citep{bib:w15} to provide additional betatron tune spread. This enhancement of the tune spread, however, may come with an impact on the dynamic aperture \citep{bib:Nikos_WP2} and goes beyond the scope of this paper.

\section{CONCLUSION}

Resistive wall impedance of LHC collimators constitutes a major part of its transverse impedance at the top energy. With the present collimation system and the brighter beams foreseen by the HL-LHC project the Landau octupole current, required to stabilize impedance-driven instabilities, exceeds the capabilities of the hardware of 570~A for the most challenging operational scenarios. The collimator impedance therefore has to be reduced in order to guarantee transverse beam stability of the HL-LHC beams.

A three-stripe prototype collimator has been installed in LHC to study the effect of low impedance coatings on beam dynamics for the HL-LHC project. Its jaws are made of MoGr with two low-resistivity coating stripes: TiN and Mo, and can be moved transversely to selectively expose the beam to the chosen material. The collimator has been installed in a slot with the maximum impact of jaw resistivity on beam dynamics next to a regular collimator for performance comparison. Resistive wall tune shifts have been measured as a function of the collimator opening to assess the impedance of each material. An unprecedented tune shift resolution of the order of $10^{-5}$ has been achieved, allowing distinguishing the impedance reduction of different low-resistivity coatings.

The results show a significant reduction of the resistive wall tune shift with novel materials compared to the presently used CFC. Uncoated MoGr reduces the tune shift by a factor 2, and the largest improvement, by a factor 4, is obtained with a 5~$\mu$m Mo coating. The tune shifts for the current CFC collimator and two of the new materials: MoGr and TiN-coated MoGr, agree within 10-20\% with the predictions of the current LHC impedance model in a wide range of collimator openings, suggesting a good identification of both the geometric and the resistive wall contributions in the experiment. The Mo coating demonstrates a two times larger resistive wall tune shift than the one expected from its DC bulk resistivity. Additional studies, such as resonant wire measurements confirmed the greater than expected resistivity of the coating, which seems to be connected to its microstructure and the sputtering method used for the prototype coating. The coating procedure was later changed to address the issue.

Based on the experimental findings, we have studied numerically the effect of upgrading the highest-contributing collimators with the novel low-resistivity jaw material. Betatron cleaning secondary collimators in IR7 are responsible for nearly a half of the LHC impedance at the frequencies relevant for the single-bunch dynamics. Upgrading them with 5~$\mu$m of Mo on MoGr reduces the total machine impedance by 30\% and an additional improvement comes from upgrading the primary collimators with uncoated MoGr. For the most challenging, ultimate operational scenario and the brightest foreseen beam, BCMS, the baseline upgrade involving 9 out of 11 secondary and 2 primary betatron cleaning collimators per beam reduces the required Landau octupole current from 940~A to 620~A after accounting for all detrimental effects: uncertainty of the impedance model, long range beam-beam encounters, linear coupling, magnet imperfections, noise from the transverse feedback or other sources, and optics errors.

It should be noted that in the present study we limited ourselves to the most conservative assumptions and did not consider effects that could be beneficial for beam stability such as transverse emittance blow-up (i.e. through intra-beam scattering), intensity loss, long range beam-beam interaction (which could be beneficial or detrimental depending on the octupole polarity), or achromatic telescopic optics. With that in mind, we believe than the present operational octupole current limit of 570~A will be sufficient to ensure beam stability for HL-LHC. Further training of the magnets can also be considered, if needed.

The collimator upgrade will begin during the long shutdown in 2019-20, when the first 4 out of 11 secondary and 2 primary betatron cleaning collimators per beam will be upgraded \citep{bib:w18}. The starting subset has been chosen to maximize the impedance reduction in the most critical, horizontal plane, and is expected to provide more than a half of the total improvement: 240~A for the most critical, BCMS beam in the present baseline. Studies will continue after the restart of LHC in Run~3 to verify the performance of the upgraded collimators and further improve the accuracy of model predictions.


\section*{Acknowledgments}

This has been performed as a part of the High-Luminosity Large Hadron Collider project. It was made possible thanks to a joint effort of many CERN groups, in particular the EN/MME, EN/STI, TE/VSC groups who made it possible to complete the construction of the prototype in a timely manner for its installation in the machine. We must also acknowledge the help of Operations, Beam Instrumentation, and Radio Frequency groups for their help in making the beam-based measurements at LHC. We would like to separately thank Gianluigi Arduini, and members of the Collimation team for numerous fruitful discussions.

\newpage
\appendix

\section{\label{sec:App_procedure} Correcting for the tune drift in the beam measurement data}

\begin{figure}[h]
  \centering
  \includegraphics[width = 2.32in]{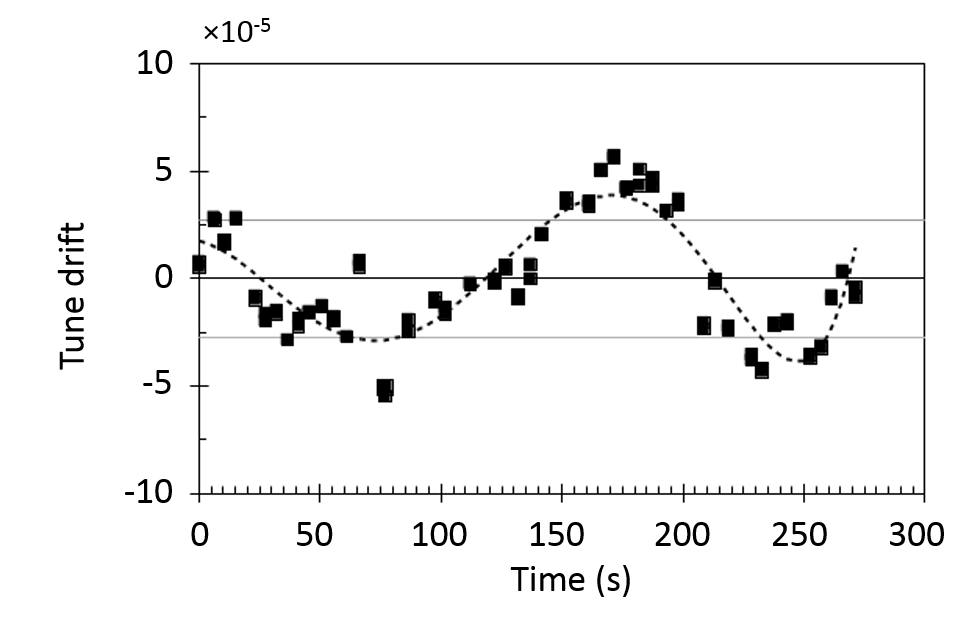}
  \caption{
  A slow tune jitter with a $\sim 100$~s period and an rms spread of (thin grey lines) is observed during the ADT excitation tune measurements. 
	}
  \label{fig:5}
\end{figure}

\begin{figure}[h]
  \centering
  \includegraphics[width = 2.32in]{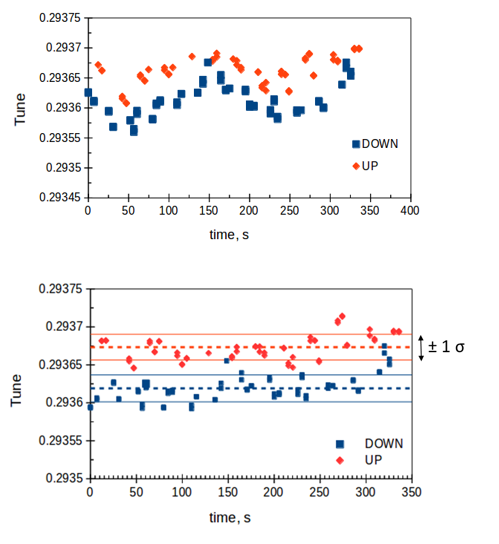}
  \caption{
  By correcting for the tune jitter one can achieve tune resolution of $\sim 10^{-5}$ and clearly distinguish the tune shift created by low impedance coatings. Tune measurements for the collimator jaws open and closed: top –- before, bottom -– after the correction. TiN stripe, $4.5\sigma$ halfgap. Solid lines represent 1 rms deviation from the mean (dashed lines).}
  \label{fig:samples}
\end{figure}

The tune drift has been removed thanks to a special measurement procedure where the collimator gaps were cycled fast between their open and closed positions while continuously exciting the beam and measuring its tune (Fig.~\ref{fig:4}). Combining the measurements at different gaps one obtains the dataset, consisting of the tune jitter (plus random errors of the measurement), which is independent of the gap. Assuming the tune drifts slowly enough, one can interpolate it with a low order polynomial and use the results to apply a correction to the measured tunes (Fig.~\ref{fig:5}). With a sufficiently large number of samples, about 100 measurements per coating stripe per collimator gap, this procedure allows resolving the individual tunes at the required $10^{-5}$ uncertainty level after correction (Fig.~\ref{fig:samples}).
 
\section{\label{sec:Geom_tune_shift} Geometric taper impedance}

\begin{figure}[h]
  \centering
  \includegraphics[width = 3.4in]{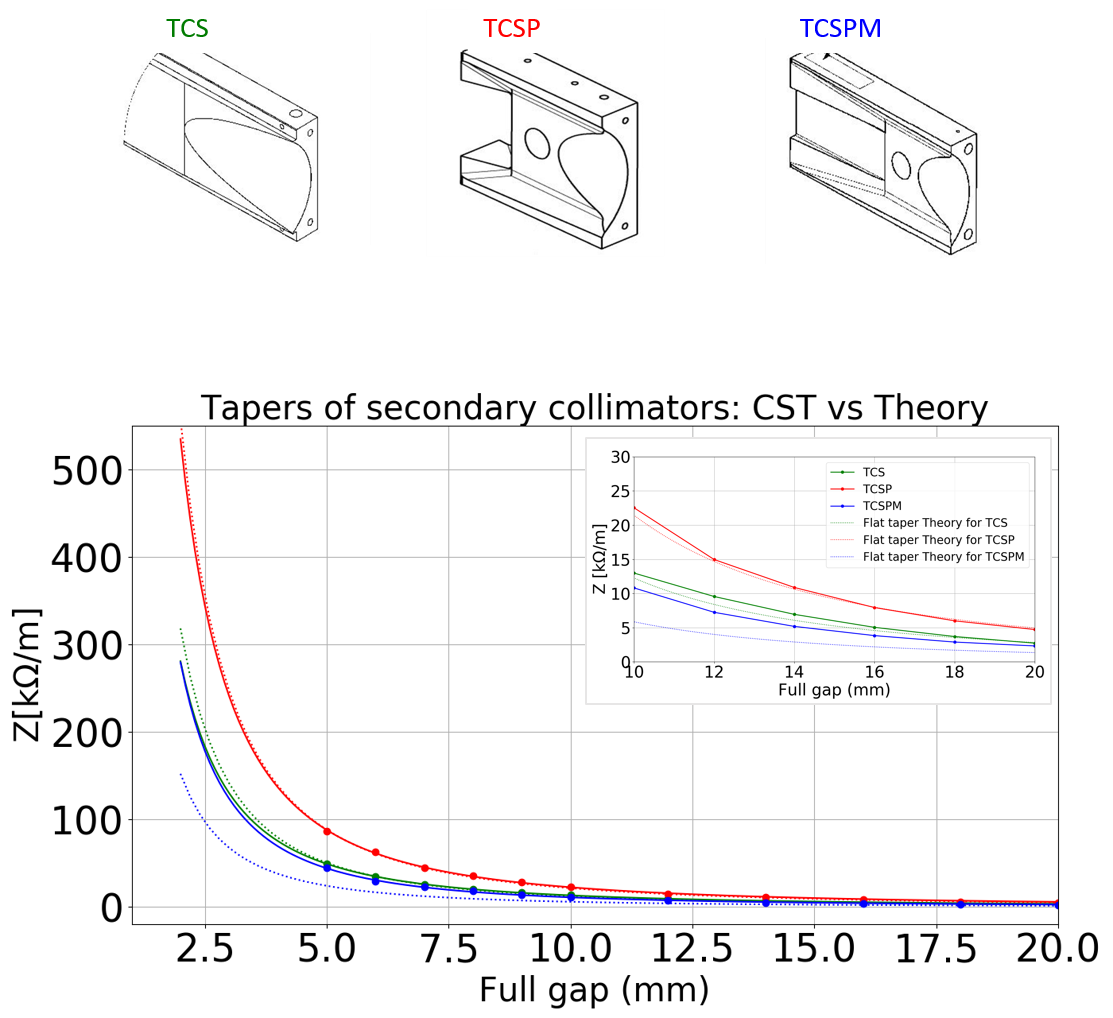}
  \caption{
  Transverse impedance as a function of half-gap in mm from CST \citep{bib:CST} simulations of the current TCSG taper (green dots), the TCSP taper (red dots), or the TCSPM taper (blue dots) compared to the flat taper theory \citep{bib:th12} used for the model (black dashed line); solid lines represent extrapolation of simulation data toward small gap heights, where numerical simulation becomes computationally intensive. Subplot in the top right corner focuses on the difference between the model and the simulation results at large gaps.}
  \label{fig:12}
\end{figure}

Different types of tapers can have drastically different geometric impedances. HL-LHC secondary collimators feature three distinct taper geometries: TCS -- the most common one presently in the machine; TCSP -- an upgraded geometry with an integrated BPM, installed on several collimators; and TCSPM -- a longer transition featuring a BPM and optimized for impedance reduction \citep{Frasciello:2014zda}, the choice for the devices to be installed in the framework of the collimator upgrade (Fig.~\ref{fig:12}, top). While the flat taper model is in good agreement with simulation for present LHC TCS tapers, it may be underestimating the impedance of TCSPM tapers by nearly a factor two (Fig.~\ref{fig:12}, bottom). Thus in order to make accurate stability predictions all existing taper geometries were numerically modelled in CST software \citep{bib:CST}. Thanks to the small share of the geometric impedance in the overall impedance of the ring, the impact of the real taper geometries turned out to be minor, at the percent level \citep{bib:Emanuela_thesis}.

\end{document}